\documentclass[pra,aps,twocolumn,nopacs,nofootinbib]{revtex4}
\usepackage{graphicx}
\usepackage{dcolumn}
\usepackage{bm}
\usepackage{amsmath}
\usepackage{epsfig}

\begin{document}
\title{Plasmons Driven by Single Electrons in Graphene Nanoislands}
\author{Alejandro~Manjavacas}
\affiliation{IQFR - CSIC, Serrano 119, 28006 Madrid, Spain}
\author{Sukosin~Thongrattanasiri}
\affiliation{IQFR - CSIC, Serrano 119, 28006 Madrid, Spain}
\author{F.~Javier~Garc\'{\i}a~de~Abajo}
\email{J.G.deAbajo@nanophotonics.es}
\affiliation{IQFR - CSIC, Serrano 119, 28006 Madrid, Spain}

\begin{abstract}
Plasmons produce large confinement and enhancement of light that enable applications as varied as cancer therapy and catalysis. Adding to these appealing properties, graphene has emerged as a robust, electrically tunable material exhibiting plasmons that strongly depend on the density of doping charges. Here we show that adding a single electron to a graphene nanoisland consisting of hundreds or thousands of atoms switches on infrared plasmons that were previously absent from the uncharged structure. Remarkably, the addition of each further electron produces a dramatic frequency shift. Plasmons in these islands are shown to be tunable down to near infrared wavelengths. These phenomena are highly sensitive to carbon edges. Specifically, armchair nanotriangles display sharp plasmons that are associated with intense near-field enhancement, as well as absorption cross-sections exceeding the geometrical area occupied by the graphene. In contrast, zigzag triangles do not support these plasmons. Our conclusions rely on realistic quantum-mechanical calculations, which are in ostensible disagreement with classical electromagnetic simulations, thus revealing the quantum nature of the plasmons. This study shows a high sensitivity of graphene nanoislands to elementary charges, therefore emphasizing their great potential for novel nano-optoelectronics applications.
\end{abstract}
\maketitle

\section{Introduction}

Light can efficiently excite plasmons (collective electron oscillations in matter), thus producing strong confinement of electromagnetic energy and huge enhancement of the associated electric fields. These phenomena have spurred a plethora of applications ranging from ultrasensitive detection \cite{NE97,KWK97,JXK05,TJO05,paper125} to cancer therapy \cite{QPA08,HSB03} and catalysis \cite{AMO01,K02}, which have been largely fueled by progress in the synthesis of noble-metal nanoparticles with increasing control over size and morphology \cite{GPM08}. Plasmons in conventional metals result from the cooperative effect of many electron-hole (e-h) virtual excitations around the Fermi level. The resulting plasmon frequencies scale with the $1/2$ power of the density of valence electrons. Understandably, massive amounts of charging are required to produce sizable frequency shifts in the plasmons \cite{CM01,HK04_2}, and therefore, their electrical control remains elusive.

This scenario has substantially changed with the arrival of graphene. Plasmons exist in this material when it is doped, but now the plasmon frequency is proportional to the $1/4$ power of the doping charge density rather than the square root of the valence electron density \cite{CGP09}. This behavior is a consequence of the peculiar electronic structure of graphene, characterized by a vanishing density of electron states at the Fermi level. Evidence for graphene plasmons has been recently reported through terahertz \cite{JGH11} and infrared (IR) \cite{FAB11,YLC12,YLZ12} optical spectroscopies, as well as through direct near-field spatial imaging \cite{CBA12,FRA12}. These studies have conclusively demonstrated that graphene plasmons can be frequency-tuned using conventional electric gating technology. In a parallel promising effort, the electro-optical tunability of graphene has also been used to modulate the plasmonic response of neighboring metallic nanoparticles \cite{ECN12,FLW12,FLW12_2}. Now, the question arises, is the singular electronic structure of extended graphene permeating the optical properties of nanometer-sized doped graphene islands as well? Can we bring the level of doping needed to sustain plasmons in a small structure down to just a single electron?

\begin{figure*}
\begin{center}
\includegraphics[width=140mm,angle=0,clip]{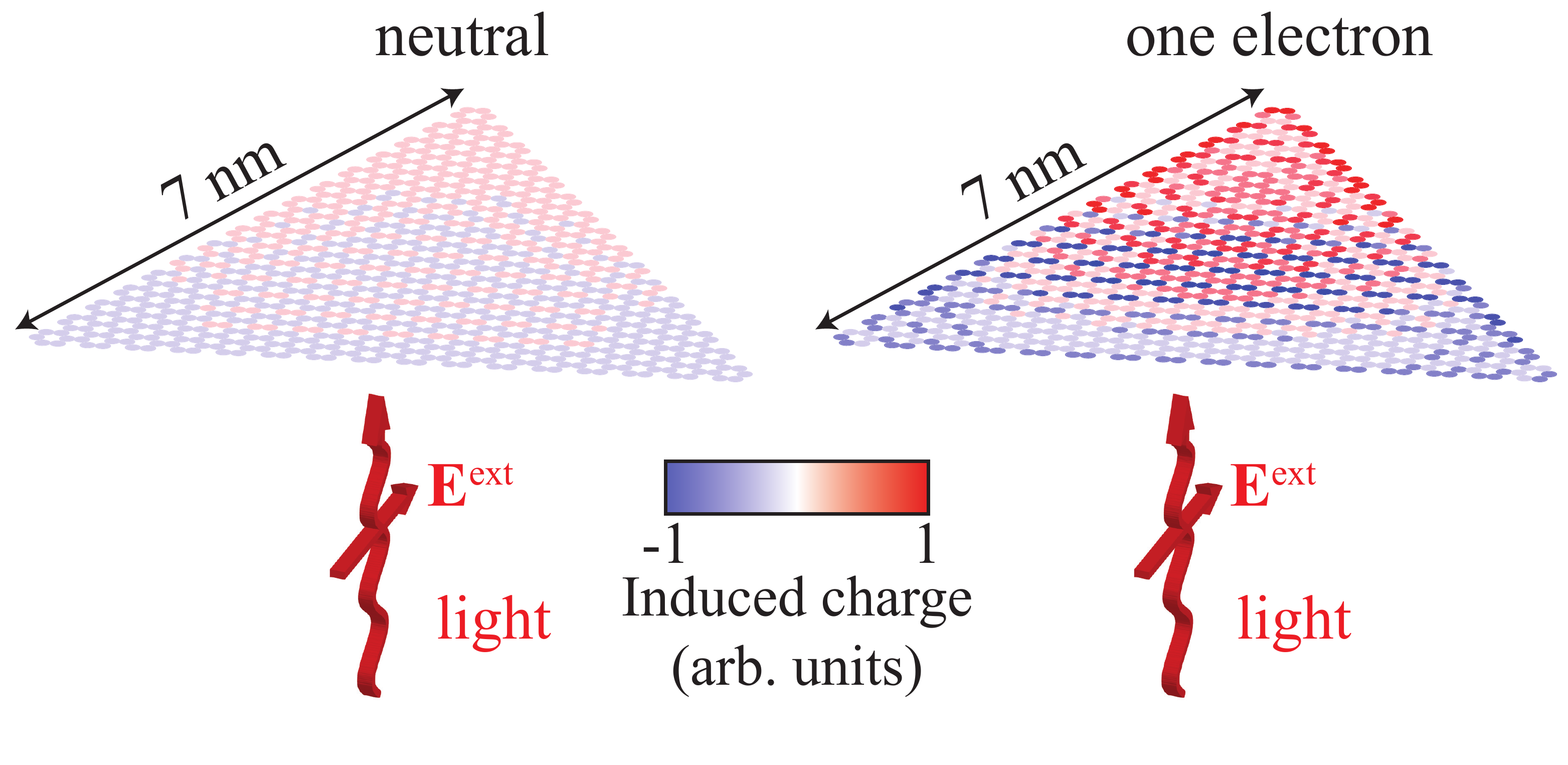}
\caption{{\bf Single-electron switching of graphene plasmons}.  A neutral graphene nanotriangle (left) shows negligible polarization under external illumination. The same nanoisland displays a 0.38\,eV plasmon resonance when it is charged with one electron, thus undergoing strong polarization (right). The density plots show the induced electron charge at the carbon sites. For comparison, the extinction cross-section is a sizable fraction of the graphene area in the resonant charged configuration and two order of magnitude smaller in the neutral island.} \label{Fig1}
\end{center}
\end{figure*}

Here we predict that IR and near-IR (NIR) plasmons in graphene nanoislands can be switched on and off by the addition or removal of a single electron. This is a remarkable property considering that the islands contain hundreds of atoms. Specifically, we focus on graphene nanotriangles, which we describe using a combination of a tight-binding model for the electronic structure and the random-phase approximation (RPA) \cite{PN1966} for the dielectric response, as explained in Appendix\ \ref{methods}. A more detailed description of the specifics of our approach is given elsewhere \cite{paper183}, essentially extending to finite graphene islands what was previously reported for more extended systems by combining tight-binding and RPA. This effort was pioneered by Wallace \cite{W1947} in graphene and graphite, and continued through outstanding analyses by other authors \cite{WSS06,BF07,CGP09}.

\section{Results and Discussion}

\begin{figure*}
\begin{center}
\includegraphics[width=160mm,angle=0,clip]{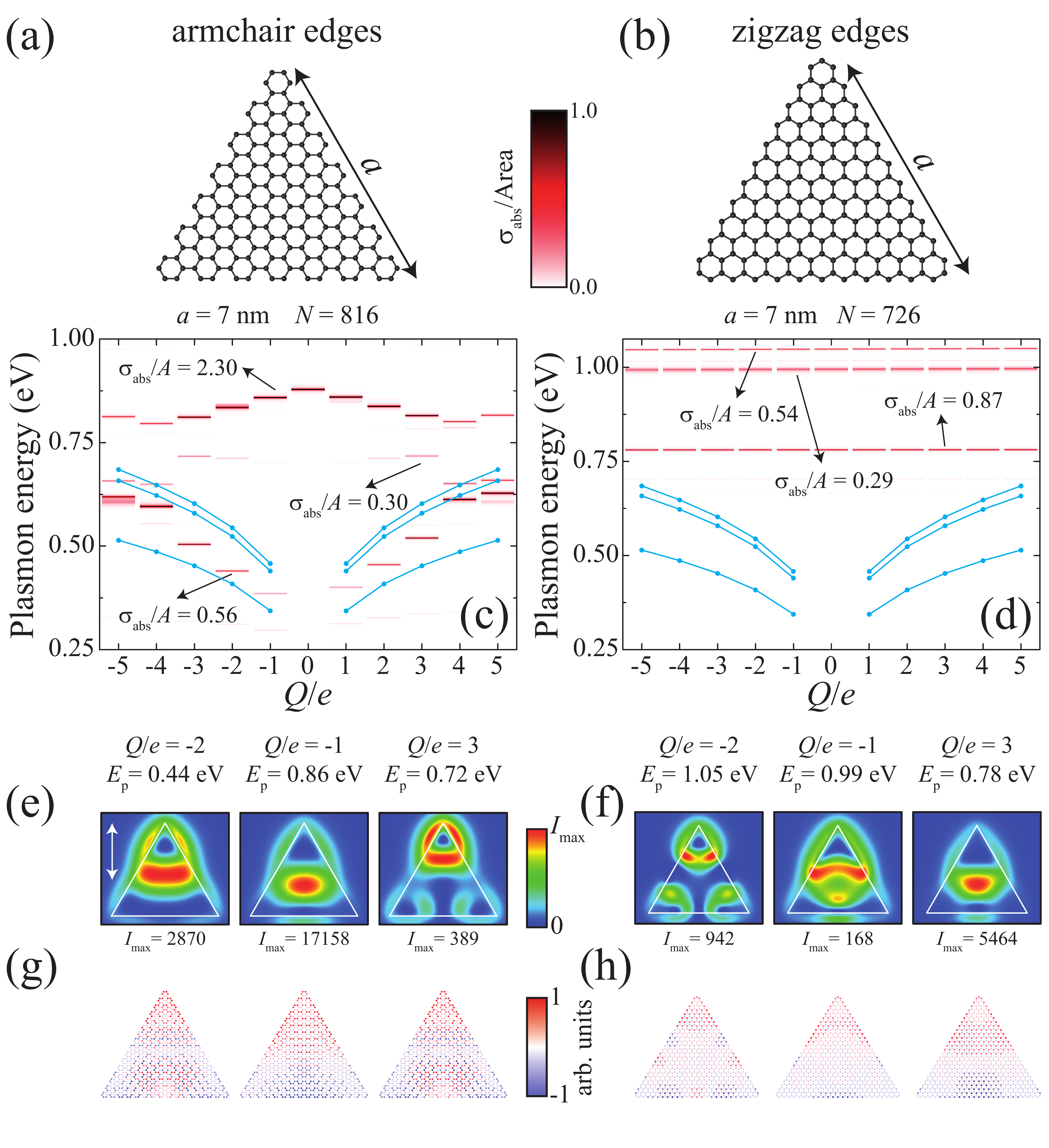}
\caption{{\bf Plasmon resonances driven by single- and few-electrons or holes in graphene triangular nanoislands}. We consider nanotriangles of either armchair (ac, {\bf a}, consisting of $N=816$ atoms) or zigzag (zz, {\bf b}, $N=726$) edges (side length $a=7\,$nm). {\bf c,} Absorption spectra of the ac island for various charge states $Q/e$, obtained from a quantum-mechanical description. Doping with electrons (holes) corresponds to $Q<0$ ($Q>0$). The absorption cross-section (color scale) is normalized to the graphene area. The plasmon energies obtained from a classical electromagnetic description of the graphene are shown by blue curves. {\bf d,} Same as {\bf c} for the zz island. {\bf e-h,} Near-field intensities ({\bf e,f}) and charge densities ({\bf g,h}) associated with selected plasmons of {c,d} for specific values of the doping charge $Q/e$ and the plasmon energy $E_{\rm p}$.} \label{Fig2}
\end{center}
\end{figure*}

Figure\ \ref{Fig1} summarizes our main finding: an electrically neutral graphene nanotriangle (with armchair edges and 7\,nm side length) does not display strong polarization when illuminated by IR light; in contrast, the addition of a single electron results in the emergence of an intense plasmon mode, so that strong polarization is produced when the light is tuned to the plasmon energy (0.38\,eV). A single electron can thus trigger the existence of plasmons in the structure. When further electrons are added to such armchair nanotriangle (Fig.\ \ref{Fig2}c), this IR plasmon undergoes a dramatic blue shift (in steps of $\sim0.1\,$eV) and it gains in strength. Notice that we denote the net charge of the structure as $Q$ in Fig.\ \ref{Fig2}, so that $Q<0$ ($Q>0$) corresponds to doping with electrons (holes).

Equally important, we find a NIR plasmon with and without doping at an energy around $\sim0.88\,$eV. In contrast to the IR plasmon, this high-energy mode undergoes redshift with increasing doping. Both the IR and the NIR plasmons give rise to large absorption cross-sections (Fig.\ \ref{Fig2}c), which can reach values exceeding the geometrical area of the graphene for large doping in the IR and for low doping in the NIR (see Appendix\ \ref{maximumabsorption}).

The plasmon shift is substantially larger than the width, thus making it clearly resolvable. Obviously, this conclusion depends on the parameter used for the intrinsic width $\hbar\tau^{-1}$ in the RPA (see Fig.\ \ref{chi0} in Appendix\ \ref{methods}). The main contributions to the width in extended graphene originate in optical-phonon losses, impurities, and disorder. These mechanisms are well described in extended graphene \cite{JBS09}, and we take $\hbar\tau^{-1}=1.6\,$meV (i.e., a dephasing time $\tau\sim400\,$fs) as a reasonable estimate for high-quality samples \cite{NGM04,NGM05}. Actually, given the small area of the islands, it should be possible to identify many of them without impurities or defects. Besides, the large shift-to-width ratios of Fig.\ \ref{Fig2} guarantee that our conclusions are still maintained with much higher losses up to the upper limit that is intuitively imposed by the lifetime of hot electrons ($\sim85\,$fs, as resolved from two-photon photoemission \cite{AGJ12}). Additionally, Landau-damping is expected to be negligible for low-energy plasmons in defect-free extended graphene \cite{PRN10}, and although plasmon decay into e-h pairs is made possible by the loss of translational symmetry in nanoislands, no substantial broadening is observed in the plasmons of armchair triangles beyond the intrinsic width $\hbar\tau^{-1}$ introduced through the RPA formula (see Fig.\ \ref{chi0} and discussion below). Actually, the plasmon energies do not overlap any intense e-h transition (see Fig.\ \ref{Fig3}c below).

Incidentally, a classical electromagnetic description of the graphene (see Appendix\ \ref{methods}) also yields IR plasmons with similar shifts as the RPA calculations (Fig.\ \ref{Fig2}a, blue curve), although the plasmon energies are substantially higher in the latter due to quantum confinement \cite{paper183}. However, the NIR plasmons and their redshifts with increasing doping are completely missed by classical theory.

\begin{figure*}
\begin{center}
\includegraphics[width=140mm,angle=0,clip]{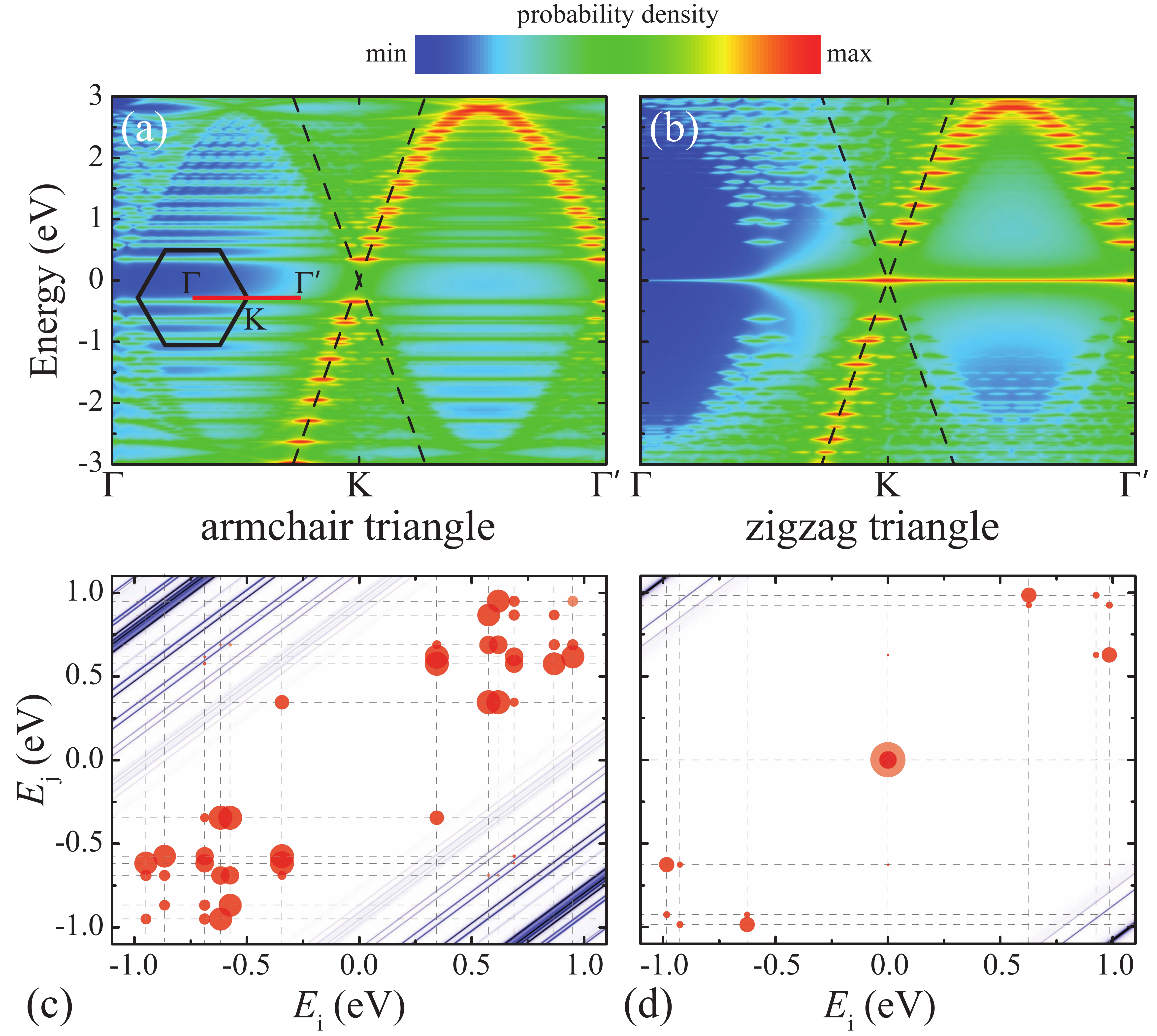}
\caption{{\bf Understanding plasmons from the electronic structure.} {\bf a,b,} Intensity of the spatial Fourier transform of the electron wave functions in the same nanoislands as in Fig.\ \ref{Fig2}. The intensity is summed over all one-electron states and it is represented as a function of energy and wave vector along the $\Gamma$K direction. The Dirac cone of extended graphene is shown by dashed lines. {\bf c,d,} Plasmons and e-h excitations mismatch, represented through the single-electron dipole-transition matrix elements $\left|\langle\psi_i|x|\psi_j\rangle\right|^2+\left|\langle\psi_i|y|\psi_j\rangle\right|^2$ as a function of initial ($E_i$) and final ($E_j$) electron-state energies. The area of the symbols is proportional to the squared matrix elements. The dark blue lines are obtained by setting $|E_i-E_j|$ to the plasmon energies of the islands under consideration (see Fig.\ \ref{Fig2}c,d).} \label{Fig3}
\end{center}
\end{figure*}

A completely different scenario is observed in zigzag nanotriangles (Fig.\ \ref{Fig2}b,d) compared to armchair structures (Fig.\ \ref{Fig2}a,c): a single plasmon appears at $\sim0.78\,$eV instead of the IR and NIR plasmons; and the plasmon energy is rather independent of doping. This seems to be connected to the presence of near-zero-energy electronic states, which are known to exist near zigzag edges \cite{FP07,E07_3,AB08_2,ZCP08,WAG10,ZCF11}. In a previous study \cite{paper183}, these states were found to produce plasmon dephasing in small islands. In the nanotriangle of Fig.\ \ref{Fig2}b, the number of such states ($24$ per electron spin, see Appendix\ \ref{effectofzigzag}) is large enough so that the addition of extra electrons does not substantially change the electronic structure. Moreover, these states do not produce dipole transitions to higher-energy states (see Fig.\ \ref{Fig3}d), and therefore, they play a dummy role in the formation of plasmons. For these reasons, mid-IR plasmons are not observed in the zigzag island, and the NIR plasmon is rather insensitive to the number of electrons added to it, until the zero-energy electronic levels are completely filled (plasmons shifts are predicted beyond this level of doping, as shown in Appendix\ \ref{effectofzigzag}).

In extended graphene, the Dirac-cone band structure leads to a gap between intra- and inter-band e-h transitions. Long-lived plasmons exist in that gap. In nanoislands, the electron parallel momentum is not a good quantum number due to the lack of translational symmetry, but we can obtain insight into the role played by the electronic structure by analyzing the spatial Fourier transform of the electron wave functions (Fig.\ \ref{Fig3}a,b). An incipient Dirac-cone structure is observed despite the finite size of the islands, which extends to a well defined Dirac cone in larger structures (see Appendix\ \ref{formationoftheDirac}). In neutral nanostructures, all states below zero energy are occupied, and remarkably, the density of states vanishes at the Dirac point in armchair islands (this is because both carbon sublattices have the same number of atoms \cite{FP07}), which explains why the addition of a few electrons causes such dramatic changes in the optical response. In contrast, zigzag islands display an intense zero-energy feature associated with edge states \cite{FP07} (see Appendix\ \ref{formationoftheDirac}), so that extra electrons do not produce significant effects in the electronic density of states, and therefore, the optical response is rather insensitive to the net charge of the structure until all zero-energy states are occupied.

The switching on of an IR plasmon with the addition of one electron poses the question, is this plasmon consisting of the oscillation of a single electron? The e-h excitation spectrum in the neutral armchair island has a gap $\sim0.7\,$eV, and the plasmon energies are actually not overlapping with those of intense e-h transitions, as we show in Fig.\ \ref{Fig3}c,d. The observed plasmons are thus involving multiple interactions among virtual e-h excitations, leading to collective electron motion. Further evidence for this is obtained by realizing that the observed plasmon energies and their characteristics are completely missed within a non-interacting electron-gas picture (see Appendix\ \ref{collectivecharacter}). Nonetheless, despite this mounting evidence of multiple e-h interactions, the role of self-screening of each electron is still a pending issue that deserves a more detailed analysis outside the scope of the present work.

Plasmon excitations in these graphene islands produce a concentration of the electromagnetic energy down to a region $\sim10^3$ smaller than the wavelength. This energy concentration leads to large levels of the field enhancement ($\sim10^3$ in intensity) upon external illumination (Fig.\ \ref{Fig2}e,f), as well as absorption cross-sections larger than the geometrical area of the graphene (see Appendix\ \ref{maximumabsorption}). The induced charge densities associated with these excitations are clearly showing dipolar excitation patterns with finer sign oscillations differentiating plasmons at different energies (see Fig.\ \ref{Fig2}g,h).

\section{Conclusion and Perspectives}

We find it remarkable that the addition or removal of a single valence electron can  switch on and off sharp plasmons in a structure already containing hundreds of valence electrons. A qualitatively similar conclusion is drawn by just examining the response of graphene nanoislands from a classical electrodynamics viewpoint. However, our quantum-mechanical simulations show that this intuition is strongly amended by the details of the atomic structure, to the point that the low-energy plasmons under consideration are simply absent from nanotriangles with zigzag terminations. In contrast, sharp IR plasmons appear in armchair-edge islands, although they are severely shifted with respect to classical theory. The predicted plasmons, and even their mere existence, are thus non-trivially  depending on the quantum mechanical properties of the underlying graphene fabric.

An important aspect of electrically tunable nanographene plasmons is that they can reach the NIR regime due to the reduced size of the structures. This is already clear in the armchair triangle of Fig.\ \ref{Fig2}a, but even higher plasmon energies reaching into the visible are obtained by reducing the size of the structure (see Appendix\ \ref{sizedependence}).

The plasmon width in armchair nanotriangles is essentially limited by the intrinsic decoherence time. Using realistic values for this parameter, we predict absorption cross-sections exceeding the geometrical area of the graphene (see Fig.\ \ref{Fig2}c and Appendix\ \ref{sizedependence}). This should allow the design of patterned graphene sheets with spacings of only a few nanometers for complete optical absorption at electrically tunable IR and NIR frequencies \cite{paper182}.

Graphene nanoislands of sizes in the range of those considered here have been fabricated with different methods \cite{LYG11,SLL12,KHK12}. However, a major challenge is the synthesis of structures with the desired edge terminations. Although armchair edges are more energetically favorable and they are observed in experiments \cite{KFE05,TCW11}, zigzag edges grow faster and dominate uncontrolled growth \cite{GME09,SCT12}. Here, we are predicting appealing optical properties for armchair nanoislands, which are nonetheless expected to grow stably under very low or very high hydrogen chemical potential \cite{GS10}. In this regard, a chemical-synthesis bottom-up approach can be beneficial to produce nanometer-sized graphene structures with well controlled edges \cite{WPM07,FLP08}.

Our prediction of single-electron-induced extreme plasmon shifts relative to the plasmon widths opens new possibilities for ultrasmall sensors based upon the observation of minute amounts of charge transfer. For example, single-molecule detection should be possible by measuring the energy shift associated with the transferred charge upon absorption of donor or acceptor molecules, thus suggesting an alternative optical approach to electrical single-molecule detection in graphene \cite{SGM07}. However, giving the large mismatch between the wavelengths of graphene plasmons and light, plasmon readout of individual nanoislands is currently a challenge. In this respect, electrical \cite{BBN11} or electron-beam plasmon excitation and detection are promising options. In particular, electron beams can maximally couple to the plasmons of graphene. The large wavelength mismatch also leads to unprecedentedly high values of the Purcell factor (quality factor divided by plasmon mode volume in units of the cubed wavelength), the field enhancement (with potential application to surface-enhanced IR absorption \cite{KLN08}), and the local density of optical states (reaching the strong light-matter interaction regime \cite{paper176}). Tightly bound plasmons in graphene nanoislands are thus a promising tool for the investigation of fundamental optical phenomena and for applications to sensing and opto-electronics.

\section*{ACKNOWLEDGMENT}

This work has been supported in part by the Spanish MICINN (MAT2010-14885 and Consolider NanoLight.es) and the European Commission (FP7-ICT-2009-4-248909-LIMA and FP7-ICT-2009-4-248855-N4E). A.M. acknowledges financial support through FPU from the Spanish ME.

\appendix

\section{Calculation methods}
\label{methods}

\subsection{Quantum-mechanical plasmon calculations}

The linear optical response of graphene nanoislands is described in the RPA \cite{PN1966}, using tight-binding electron wave functions as input \cite{W1947}. For the low photon energies under consideration, only $\pi$ valence electrons contribute to the response. The electron wave functions are expressed in a basis set consisting of one $2p$ orbital per carbon site (denoted $|l\rangle$), oriented perpendicularly with respect to the graphene plane and occupied with one electron on average in undoped structures. A tight-binding Hamiltonian is formulated in such basis set, with nonzero elements connecting nearest neighbors through a $-2.8\,$eV hopping energy \cite{CGP09}. One-electron states $\sum_l a_{jl}|l\rangle$ and energies $\varepsilon_j$ are obtained upon diagonalization of this Hamiltonian, yielding the well-known Dirac-cone electronic structure of graphene \cite{CGP09,BOS07}. Here, $a_{jl}$ is the amplitude of orbital $|l\rangle$ in state $j$. For the graphene islands and charge states under consideration, electron states are filled up to affordable values of the Fermi energy $E_F<\,$eV relative to the Dirac point. The site-dependent induced charge is then obtained from the self-consistent potential $\phi$ as
\begin{equation}
\rho_l=\sum_{l'}\chi_{ll'}^0\phi_{l'}, \label{rhol}
\end{equation}
where
\begin{equation}
\chi_{ll'}^0(\omega)=\frac{2e^2}{\hbar}\sum_{jj'}\left(f_{j'}-f_j\right)\frac{a_{jl}a_{jl'}^*a_{j'l}^*a_{j'l'}}{\omega-(\varepsilon_j-\varepsilon_{j'})+i/2\tau}
\label{chi0}
\end{equation}
is the non-interacting RPA susceptibility, $\omega$ is the light frequency, $f_j=\{\exp[(\hbar\varepsilon_j-E_F)/k_BT]+1\}^{-1}$ is the occupation fraction of state $j$ at temperature $T$ (300\,K throughout this work), and $\tau$ is an intrinsic relaxation time. The latter is a critical parameter, which we fix to $\hbar\tau^{-1}=1.6\,$meV, a value compatible with estimates obtained from measured DC mobilities \cite{NGM04,NGM05,BSJ08}. The self-consistent potential is obtained from the external potential $\phi^{\rm ext}$ and the induced charges as
\begin{equation}
\phi_l=\phi^{\rm ext}_l+\sum_{l'}v_{ll'}\rho_{l'}, \label{phil}
\end{equation}
where $v_{ll'}$ is the Coulomb interaction between electrons at orbitals $l$ and $l'$. Finally, we calculate extinction cross-sections and near-field intensities from the self-consistent charge $\rho=[\chi^0/(1-v\cdot\chi^0)]\cdot\phi^{\rm ext}$ derived from Eqs.\ (\ref{rhol,phil}) upon illumination by an external field ${\bf E}^{\rm ext}$ (i.e., we take $\phi_l^{\rm ext}=-{\bf R}_l\cdot{\bf E}^{\rm ext}$, where ${\bf R}_l$ is the position vector of site $l$). For example, the cross section reads $\sigma^{\rm abs}=(4\pi\omega/c){\rm Im}\{\alpha\}$, where $\alpha=(1/E^{\rm ext})\sum_l x_l\rho_l$ is the nanoisland polarizability. Further computational details can be found elsewhere \cite{paper183}.

\subsection{Classical electromagnetic modeling}

We compare our quantum-mechanical results with classical finite-element (COMSOL) calculations in which the graphene is described as a thin film of thickness $t$ and dielectric function $1+4\pi i\sigma/\omega t$, where $\sigma(\omega)=(ie^2E_F/\pi\hbar^2)/(\omega+i\tau^{-1})$ is the Drude optical conductivity.

\section{Geometry of the nanoislands}

\begin{figure}
\begin{center}
\includegraphics[width=70mm,angle=0,clip]{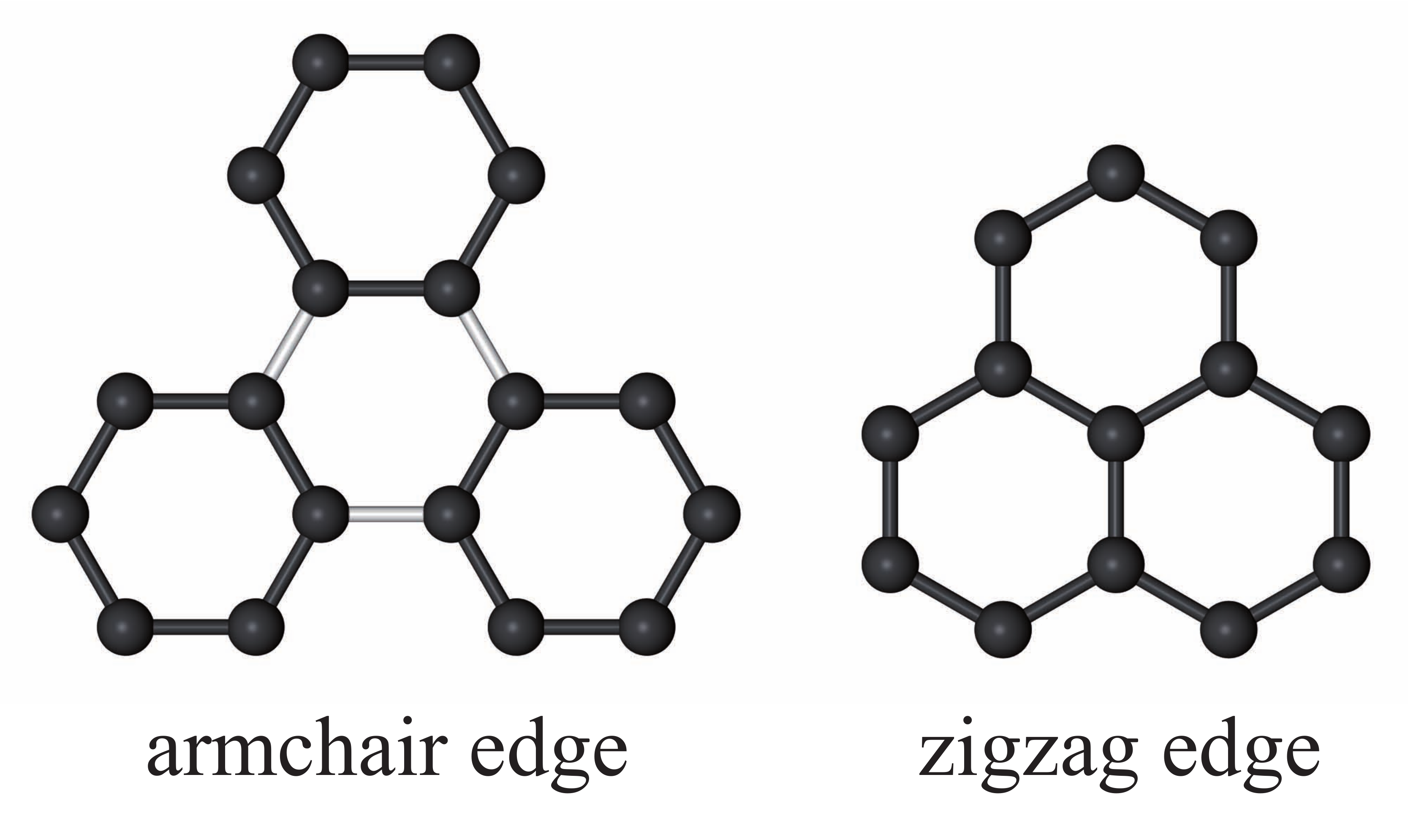}
\caption{{\bf Definition of $n$ in armchair and zigzag nanotriangles.} We define $n$ as the number of hexagons along the side ($n=2$ in these examples).} \label{N}
\end{center}
\end{figure}

Most of the calculations reported in this paper refer to graphene nanotriangles of either armchair or zigzag edges centered around the middle point of a carbon hexagon. We consider the edge atoms to be passivated with hydrogen, and therefore, we treat them as the rest of the atoms in the structure, with the same hopping to their nearest neighbors. The number of atoms in the nanotriangles is $3n(n+1)$ for armchair edges and $(n+2)^2-3$ for zigzag edges, where $n$ is the number of hexagons along the side. For example, the $n=2$ triangles of Fig.\ \ref{N} contain 18 and 13 carbon atoms, respectively. The side length of the triangles is $a=\left(3n-1\right)a_0$ for armchair edges and $a=\sqrt{3}na_0$ for zigzag edges, where $a_0=0.142$\,nm is the nearest-neighbor distance.

\section{Size dependence}
\label{sizedependence}

\begin{figure*}
\begin{center}
\includegraphics[width=160mm,angle=0,clip]{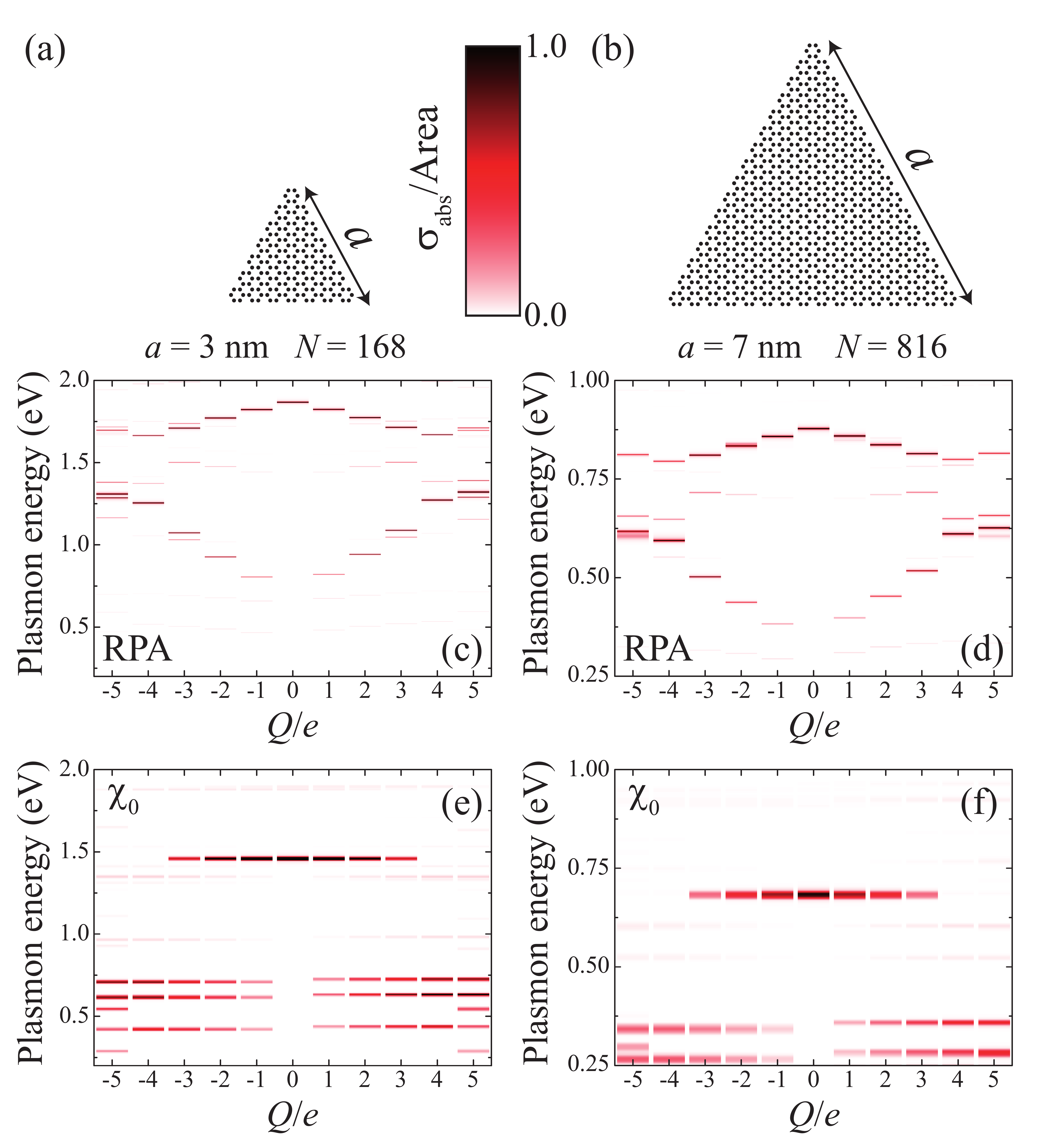}
\caption{{\bf Size dependence and collective character of nanotriangle plasmons.} We study plasmons in a graphene armchair nanotriangle of side $a=3\,$nm (left), compared with those of the $a=7\,$nm island considered in the this paper and reproduced here for convenience (right). We show the carbon atomic structure of the nanotriangles in (a) and (b), respectively. The absorption cross-section normalized to the graphene area is calculated for these two nanotriangles from the self-consistent RPA ((c) and (d)), compared with the non-self-consistent RPA ((e) and (f)). The spectral dependence of the cross section is given as a function of the number of elementary charges in the nanoislands, $Q/e$.} \label{small}
\end{center}
\end{figure*}

We show in Fig.\ \ref{small} extinction spectra for armchair nanotriangles of two different sizes. The main effect of reducing the size of the graphene island is an increase in the plasmon energies, which is significantly larger than the $1/\sqrt{a}$ dependence on side length $a$ predicted by a classical electromagnetism description \cite{paper181}. However, like in classical theory, the absorption cross-section produced by these plasmons is roughly proportional to the area.

\section{Collective character of the plasmons}
\label{collectivecharacter}

In the random-phase approximation (RPA) here used to describe the optical response of graphene nanoislands, the induced density is writen as $\rho=[\chi^0/(1-v\cdot\chi^0)]\cdot\phi^{\rm ext}$ in terms of the external potential $\phi^{\rm ext}$, the Coulomb interaction between atomic sites $v$, and the noninteracting RPA susceptibility \cite{paper183} (see Appendix\ \ref{methods}). The collective character of the plasmons is clearly captured in a multiple-scattering fashion by the use of the self-consistent potential $[1/(1-v\cdot\chi^0)]\cdot\phi^{\rm ext}$ in the above expression. We compare in Fig.\ \ref{small} these full RPA calculations (Fig.\ \ref{small}c,d) with those obtained by removing the self-consistency and calculating the induced charge as $\rho=\chi^0\cdot\phi^{\rm ext}$ (Fig.\ \ref{small}e,f). This leads to dramatic changes in the plasmon spectrum: the plasmon dispersion with varying doping charge is removed, and the spectral features are now peaked at the positions of dominant electron-hole-pair transitions (see Fig.\ \ref{Fig3}c). This is in contrast to the plasmons obtained from self-consistent calculations, the energies of which depend on the interaction among different electron-hole-pair virtual transitions, as a manifestation of the collective electron motion involved in these excitations.

\begin{figure*}
\begin{center}
\includegraphics[width=160mm,angle=0,clip]{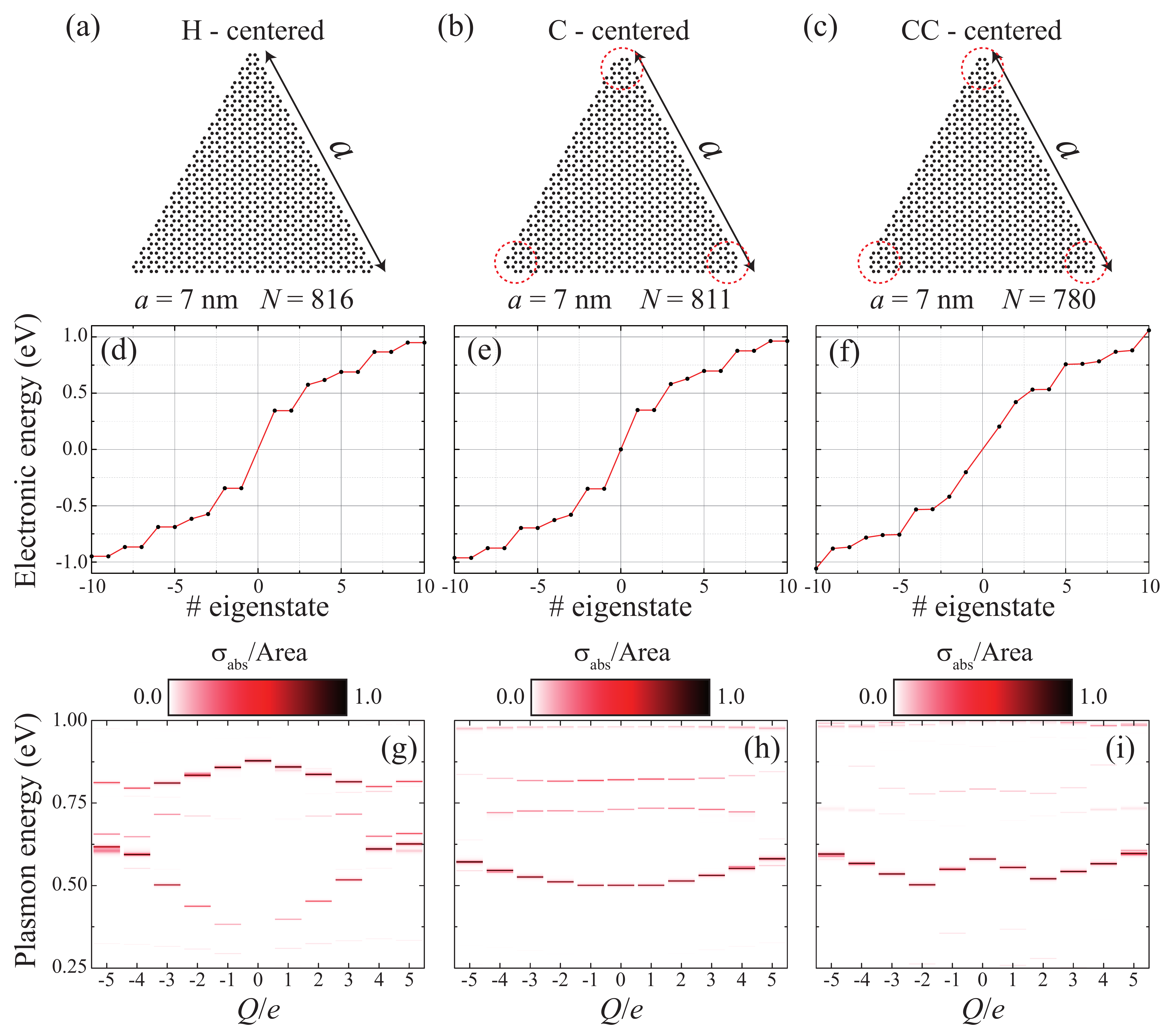}
\caption{{\bf Dependence on the choice of triangle center.} Atomic structure (a-c), states near the Dirac point (d-f), and extinction spectra (g-i) of armchair nanotriangles with $a=7\,$nm side length centered at the center of a carbon hexagon (H-centered) (a,d,g), at a carbon atom (C-centered) (b,e,h), or at the center of a nearest-neighbor carbon bond (C-C-centered) (c,f,i).} \label{HCC}
\end{center}
\end{figure*}

\section{Dependence on the choice of triangle center}

Plasmons in graphene nanoislands are very sensitive to small structural details of the carbon lattice. Besides the strong dependence on the type of edges observed in Fig.\ \ref{Fig2}, we show in Fig.\ \ref{HCC} a large variation in the plasmonic spectrum with the position of the triangle center. In Fig.\ \ref{HCC}a,d,g and in the rest of the figures of this work, we present calculations obtained with the center located at the middle point of a carbon hexagon (H-centered). When the triangle center is chosed at a carbon site (C-centered, Fig.\ \ref{HCC}b,e,h) or at the center of a nearest-neighbors bond (C-C-centered, Fig.\ \ref{HCC}c,f,i), the corners of the atomic structure present a certain degree of asymmetry (see Fig.\ \ref{HCC}b,c) and the one-electron energies are substantially modified (Fig.\ \ref{HCC}e,f). This leads to dramatic changes in the extinction spectra (Fig.\ \ref{HCC}h,i), which are still exhibiting a clear plasmon dispersion with varying doping charge. In particular, the lowest-energy plasmon of the C-C-centered triangle (Fig.\ \ref{HCC}i) shows a bump in energy at zero doping, in contrast to the lowest-energy plasmon of the H-centered triangle (Fig.\ \ref{HCC}g), which steadily increases in energy with doping.

\begin{figure*}
\begin{center}
\includegraphics[width=140mm,angle=0,clip]{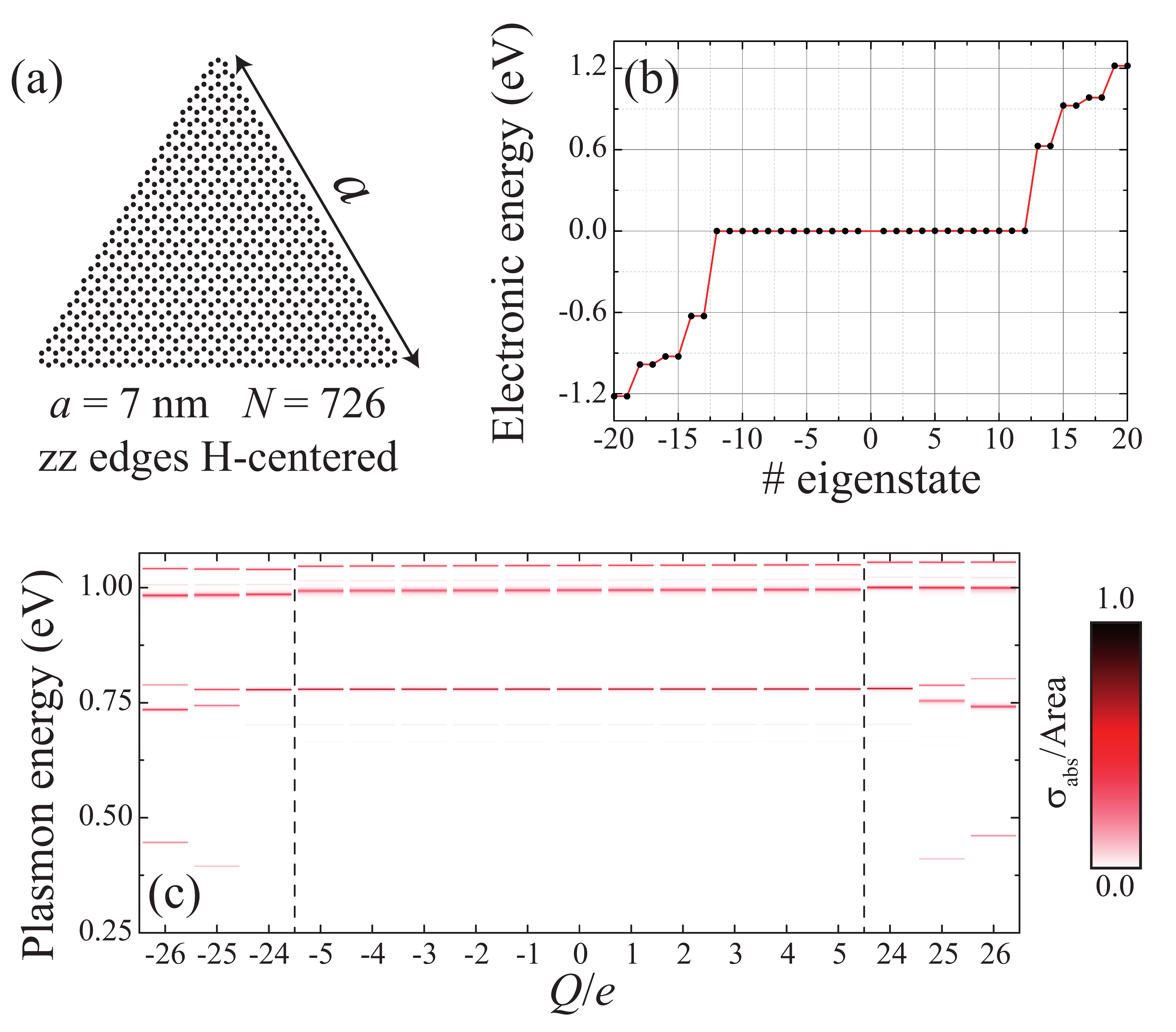}
\caption{{\bf Emergence of plasmons upon filling all zero-energy electronic states in a zigzag triangle.} The zigzag triangle represented in (a) has 24 zero-energy states per electron spin at the Dirac point (see (b)), half of which are occupied in the neutral configuration. No significant change in the plasmon energy and strength is observed by adding electrons (holes) until the zero-energy states are fully occupied (empty). This happens for $|Q|/e=24$. In contrast, the addition (removal) of each further electron beyond this point produces sizable changes, as shown in the extinction cross-section plotted in (c) for $|Q|/e>24$.} \label{zz}
\end{center}
\end{figure*}

\section{Effect of zigzag edges and zero-energy electronic states}
\label{effectofzigzag}

Zigzag edges are known to produce electron states near the Dirac point that are concentrated near the atomic edges. The number of such states is equal to the difference between the number of atoms in the two carbon sublattices of the graphene nanostructure \cite{FP07,AB08_2,WAG10}. This is the case of the C-centered triangle of Fig.\ \ref{HCC}b, which contains zigzag edges near the corners that lead to the existence of one zero-energy state. This state is half-filled in the neutral structure. Interestingly, adding or removing a single electron to the nanotriangle does not produce any significant change in the plasmonic spectrum because it ends up in an edge-localized electronic state that cannot undergo large dipole transitions. However, the addition or removal of two electrons causes a sizeable shift. One can therefore speculate that electrons added (removed) to (from) the zero-energy states have no effect in the collective plasmon excitations. Another piece of evidence pointing to this direction is provided by the zigzag triangle considered in Fig.\ \ref{Fig3}b,d,f, which contains a large number of zero-energy states and does not suffer any plasmon shifts with the addition of a small number of electrons or holes. In order to test this hypothesis, we consider the same zigzag triangle in Fig.\ \ref{zz}, subject now to much higher doping. Interestingly, the plasmonic spectrum remains nearly unchanged while electrons or holes are added to the nanoisland. However, when the zero-energy states are completely filled (depleted) and further electrons (holes) are added to the structure, the plasmons undergo strong variations in energy with every new electron (hole). This is strong evidence that zero-energy states do not contribute to plasmons, and the tunability observed in graphene islands with varying doping is quenched by the presence of these states. In summary, zigzag edges produce a net difference in the number of atoms in the two carbon sublattice of graphene nanoislands, leading to the presence of zero-energy states localized near the edges, the optical excitation of which is rather inefficient because they cannot undergo large dipole transitions, and therefore, zigzag edges are detrimental to the tunability of graphene plasmons via doping.

\begin{figure*}
\begin{center}
\includegraphics[width=160mm,angle=0,clip]{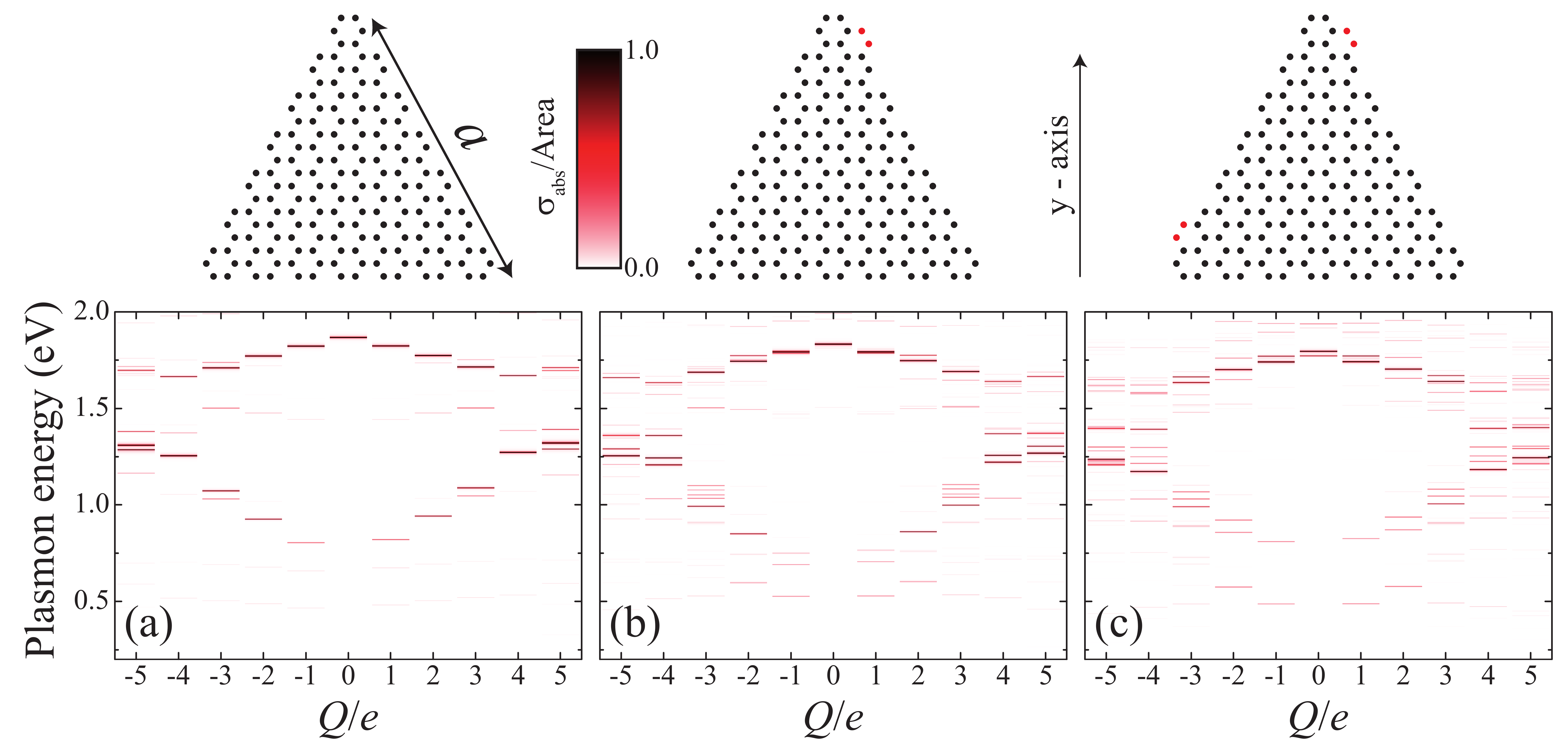}
\caption{{\bf Effect of disorder in the edges.} Spectral changes in doped armchair nanotriangles when disorder is introduced in the edges. The spectral features of a perfect nanotriangles of side length $a=3\,$nm (a) are increasingly split as additional atoms are added to the edges: 2 atoms in (b) and 4 atoms in (c).} \label{disorder}
\end{center}
\end{figure*}

\section{Effect of disorder in the edges}

The robustness of the spectra of armchair nanotriangles is tested in Fig.\ \ref{disorder} against the degree of disorder in the edges. In particular, this figure shows that the spectral features of an immaculate doped $a=3\,$nm triangle are increasingly split as extra atoms are added to the edges. Additionally, low-energy features are becoming more intense. Incidentally, the number of atoms in both carbon sublattices remains the same in all the structures considered in Fig.\ \ref{disorder} (i.e., the two atoms added in each step belong to different sublattices), so that there are neither zero-energy states nor effects associated with dummy zero-energy electrons, as described in the previous section.

\begin{figure*}
\begin{center}
\includegraphics[width=160mm,angle=0,clip]{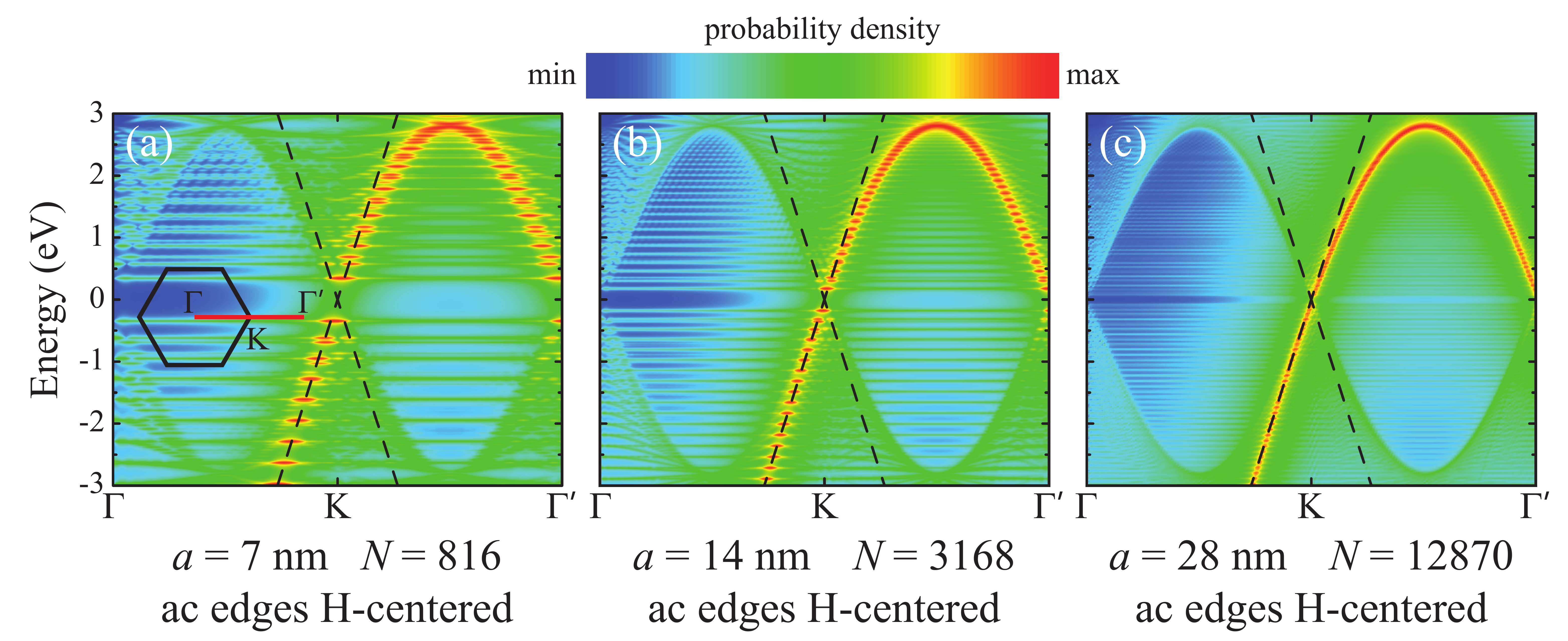}
\caption{{\bf Building up the Dirac cone structure.} Intensity of the Fourier transform of all one-electron wave functions in armchair nanotriangles of increasing size. This is the continuation of Fig.\ \ref{Fig3}a towards larger sizes of the nanoisland.} \label{Fourier}
\end{center}
\end{figure*}

\section{Formation of the Dirac cone in large islands}
\label{formationoftheDirac}

We show in Fig.\ \ref{Fourier} the formation of the Dirac cone structure as the size of the nanotriangle is increased. The Dirac cone emerges in the momentum-energy representation of the electronic bands in extended graphene. However, the momentum is not a good quantum number in finite nanoislands due to the lack of translational symmetry. Instead, we have taken the spatial Fourier transform of each and all of the one-electron states and represented the sum of their intensities as a function of wave vector. This produces an energy-momentum structure that is reminiscent of the Dirac cone, which is more clearly emerging as the size of the nanotriangles increases.

\begin{figure}
\begin{center}
\includegraphics[width=80mm,angle=0,clip]{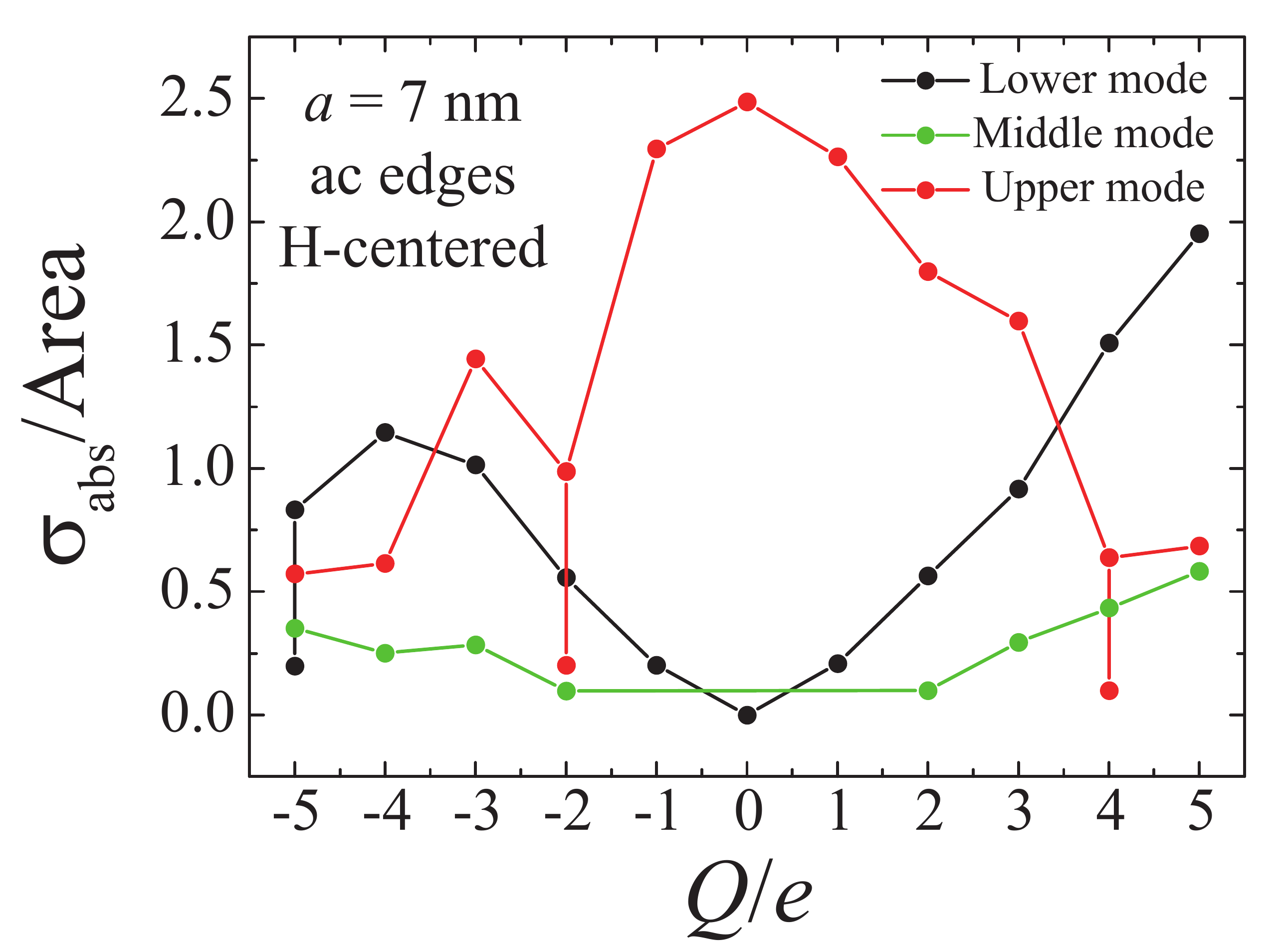}
\caption{{\bf Absorption cross-sections exceeding the graphene geometrical area.} Peak absorption cross-sections associated with the plasmons of Fig.\ \ref{Fig2}c.} \label{sigma}
\end{center}
\end{figure}

\section{Maximum absorption cross-section}
\label{maximumabsorption}

The absorption cross-section at the peak of the features associated with the excitation of plasmons can exceed the geometrical area of the graphene. We show this in Fig.\ \ref{sigma} for the plasmons of the island considered in Fig.\ \ref{Fig2}a. These cross-sections are thus large enough to produce complete optical absorption in a surface decorated with nanoislands \cite{paper182}. Importantly, the maximum cross-section is directly proportional to the intrinsic decoherence time $\tau$, which can be increased due to thermal and disorder effects \cite{JBS09}. Nonetheless, our choice of $\hbar\tau^{-1}=1.6\,$meV is compatible with estimates based upon reported DC mobilities for high-quality graphene \cite{JBS09}.


\end{document}